\documentclass[
    ,final            
  ]
  {aipproc}

\layoutstyle{6x9}

\begin{document}

\title{INTEGRAL Spectrometer Analysis of GRB030227 \& GRB030131\\
}

\author{L. Moran}{
  address={Department of Experimental Physics, University
  College Dublin, Ireland}
}

\author{L. Hanlon}{
  address={Department of Experimental Physics, University
  College Dublin, Ireland}
}

\author{B. McBreen}{
  address={Department of Experimental Physics, University
  College Dublin, Ireland}
}

\author{R. Preece}{
 address={Department of Physics, University of Alabama at
Huntsville, USA}
}

\author{Y. Kaneko}{
 address={Department of Physics, University of Alabama at
Huntsville, USA}
}

\author{O.R. Williams}{
 address={Science Operations and Data Systems Division of
  ESA/ESTEC, SCI-SDG, NL-2200 AG Noordwijk, The Netherlands}
}

\author{K. Bennett}{
 address={Science Operations and Data Systems Division of
  ESA/ESTEC, SCI-SDG, NL-2200 AG Noordwijk, The Netherlands}
}

\author{R. Marc Kippen}{
 address={Space and Remote Sensing Sciences, Los Alamos National
Laboratory, USA}
}

\author{A. Von Kienlin}{
 address={Max-Planck-Institut f\"{u}r Extraterrestrische Physik, 85748
  Garching, Germany}
}

\author{V. Beckmann}{
 address={NASA Goddard Space Flight Center, University of Maryland
 Baltimore County, USA}
}

\author{S. McBreen}{
  address={Department of Experimental Physics, University
  College Dublin, Ireland}
}

\author{J. French}{
  address={Department of Experimental Physics, University
  College Dublin, Ireland}
}

\begin{abstract}
The spectrometer SPI on board INTEGRAL is capable of high-resolution
spectroscopic studies in the energy range 20\,keV to 8\,MeV for GRBs which
occur within the fully coded field of view (16$^\circ$ corner to
corner). Six GRBs occurred within the SPI field of view between
October 2002 and November 2003. We present results of the analysis of
the first two GRBs detected by SPI after the payload performance and
verification phase of INTEGRAL. 
\end{abstract}

\vspace{-.4cm}

\maketitle

\vspace{-.4cm}


\vspace{-.4cm}

\section{Introduction}

\vspace{-.2cm}

On October 17$^{\rm th}$ 2002, ESA's gamma-ray observatory INTEGRAL was
successfully launched from the Ba\"{i}konur Cosmodrome in
Kazakhstan. INTEGRAL has a burst alert system (IBAS) which carries out
rapid localisations for gamma-ray bursts (GRBs) incident on the IBIS
detector. These co-ordinates are then distributed, allowing for fast
follow up observations at other wavelengths \cite{mere:2003}. The
main instruments IBIS and SPI also contribute greatly to INTEGRAL's
GRB capabilities. IBIS is a high resolution imager \cite{uber:2003}
with angular resolution of 12\,arcminutes for sources within its
9$^\circ$ $\times$ 9$^\circ$ fully coded field of view and broadband spectral
capabilities.  SPI is optimised for spectroscopic study of gamma-ray
sources, with some imaging capabilities. IBIS and SPI are complemented
by two smaller instruments, JEM-X and OMC, which monitor gamma-ray
sources at x-ray and optical wavelengths. 

SPI consists of 19 high purity Germanium detectors actively cooled to
a temperature of $\sim$\,85\,K to provide an energy resolution FWHM of
2.5\,keV at 1\,MeV \cite{knod:2002} in the range
20\,keV-8\,MeV. SPI's imaging capabilities are due to a coded mask
comprising of 127 tungsten elements, with a thickness of 30\,mm,
placed at a distance of $\sim$\,1.7\,m from the detection plane,
providing an angular resolution of 2.8$^\circ$. The good angular
resolution combined with excellent spectral resolution make SPI an
ideal instrument for spectral studies of the prompt emission of GRBs.

GRBs, first detected over 35 years ago, are an intriguing phenomenon
and remain at the forefront of research in astrophysics. The discovery
by BeppoSAX of afterglows in the x-ray \cite{costa:1997} and subsequent discoveries at
optical \cite{vanp:1997} and radio \cite{frail:1997} wavelengths have led to redshift
measurements \cite{metz:1997} for $\sim$\,40 bursts ranging from z\,=\,0.168\,-\,4.5. A
theory of gamma-ray bursts must provide a mechanism capable of a non-thermal
energy output of the order of 10$^{52}$-10$^{54}$\,ergs by compact
sources at cosmological distances.

With a large detector area of approximately 500\,cm$^2$ and
its high spectral resolution, SPI can address the long-standing
controversy over the existence of short-lived spectral features in
GRB spectra, previously searched for with varying degrees
of success \cite{mura:1988,briggs:1997}. A confirmation of line
features and details of specific spectral features could contribute
to the debate on the connection between GRBs and core-collapse
supernovae \cite{hjo:2003}. In addition, the broad energy coverage of SPI
(20\,keV-8\,MeV) is well suited to constrain the spectral shape, both
below and above the energy at which the GRB power output is typically
peaked ($\sim$\,250\,keV). Study of the spectral shape of the prompt
emission of a GRB at the onset of the afterglow from SPI data could
reveal the activity of the central engine that leads to the production
of an afterglow. At the high energy end, there may exist a hard
spectral upturn as recently found by Gonz\'{a}lez et al. (2003)
\nocite{gon:2003} in archival CGRO data of GRB941017. 

\vspace{-.4cm}

\section{SPI Data Analysis of GRBs}

\vspace{-.2cm}

A standard SPI pointing has a duration of $\sim$\,35\,minutes. In this
Science Window (ScW) the numbers of single, double and higher
multiplicity events striking each detector are recorded according to
the energy (16384 channels). Photon by photon information, which
contains details of the detector struck, the energy deposited and the
exact time, is available for multiple events and all events
analysed by the Pulse Shape Discriminator (PSD), about 10\% of the
total. The standard SPI pipeline is designed to process sets of ScWs
\cite{couv:2003}, whereas a GRB lasts only a fraction of this
duration. Therefore, a modified analysis procedure (see Fig. 1) is
used in which the start and end times and the best known position of
the GRB are manually inserted.

\begin{figure}[h]
  \includegraphics[width=\textwidth]{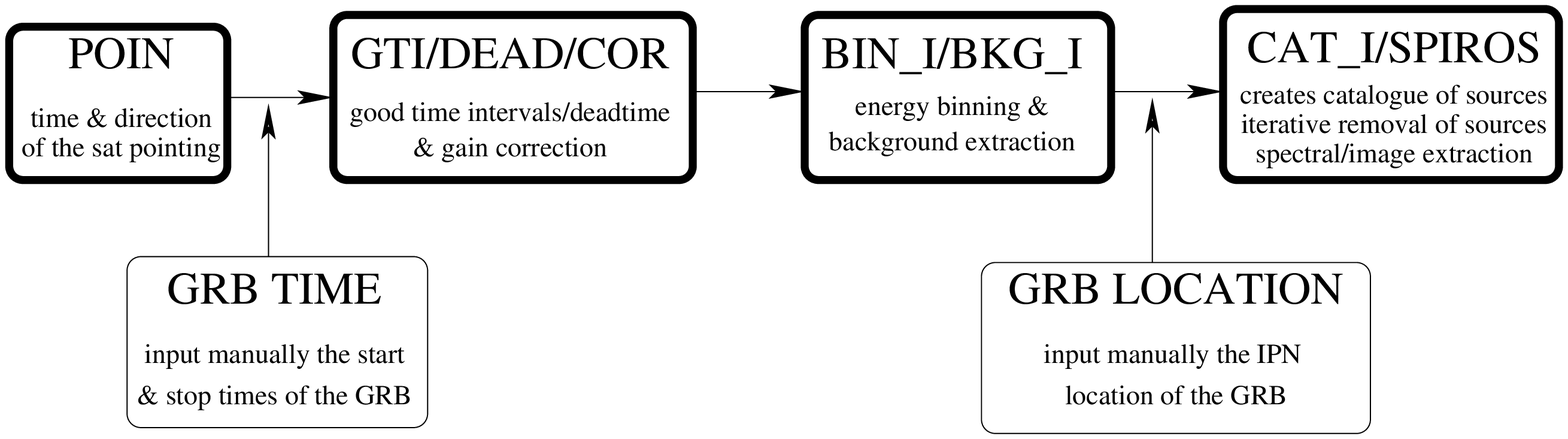}
  \caption{Flow chart of the SPI instrument specific software pipline,
  including the additional steps necessary for GRB analysis.}
\end{figure}

\vspace{-.2cm}

\section{Results}

\vspace{-.2cm}


{\bf GRB030131} At 07:39:00 UT on January 31$^{\rm st}$ 2003, SPI detected a
gamma-ray burst of duration $\sim$\,60\,s, the first since commencing full
operational status. The corrected INTEGRAL position of the burst was
given in {\it GCN 1847} as RA 202.13$^\circ$ and DEC 30.68$^\circ$ 
with a 5' radius error. The analysis of GRB030131 is complicated by
the fact that 10\,s after the GRB onset, the spacecraft started
slewing and the remaining 50\,s of the burst, including the brightest
portion, occurred during this manoeuvre. 

There are two main problems which arise when a SPI observation takes
place during a satellite slew. The first is that not all types of
events are recorded during a slew. To facilitate compression and
transmission of the data from the preceding steady pointing, only
multiple events and those single  events analysed by the PSD are
downlinked, together with the technical- and science-housekeeping data,
with the loss of $\sim$90\% of the single events.

The second complication is that to utilise the analysis procedure
 developed, the pointing  direction of the instrument and  details of
 the slew must be established and manually inserted. In particular,
 the RA and DEC of the satellite's orientation need to be chosen to
 reflect the motion of the spacecraft. As burst emission is evident
 for one third of the length of the slew, the co-ordinates of the SPI
 field of view were deemed to lie one sixth of the angular distance
 from the preceding steady pointing to the subsequent one. 

Spectral anaysis was not possible with the limited telemetry received, but
 using the imaging capabilities of SPI the burst was located with a
 detection significance of 3.6\,$\sigma$. The first row in {\it Table 1} gives
 the results obtained using the modified pipeline for the brightest
 10\,s of the burst.

\vspace{.2cm}

\begin{table}[h]
\begin{tabular}{cccccc}
\hline
   \tablehead{1}{r}{b}{RA}
  & \tablehead{1}{r}{b}{DEC}
  & \tablehead{1}{r}{b}{$\sigma$}
  & \tablehead{1}{r}{b}{$Flux \, (ph/cm^2/s)$}
  & \tablehead{1}{r}{b}{$Energy\, range\, (keV)$}   \\
\hline
200.650 & 30.210 & 3.6 & 0.86 $\pm$ 0.24 & 20-500\\
201.967 & 31.117 & 7.0 & -  & 20-8000\\
\hline
\end{tabular}
\caption{Results for the brightest 10\,s of GRB030131.
}
\label{tab:b}
\end{table}

Taking into account the SPI localisation precision of
2.8$^\circ$, the SPI and IBIS locations are in agreement. The
flux obtained is also consistent with that derived from IBIS
data \cite{got:2003}. The second row in {\it Table 1} shows the
results of analysis performed using the science-housekeeping
data. This method does not require an input location and yet still
locates a source consistent with the GRB location with a significance
of 7\,$\sigma$.

{\bf GRB030227} began at 08:42:04 UT on 27$^{\rm th}$ Febuary 2003 and had a
duration of 18\,s. Though a weak burst \cite{mer:2003}, SPI images the
burst with a detection significance of 7.5\,$\sigma$ and location in agreement
with IBIS. A power law model fit to GRB030227 using XSPEC yields a
photon index of 1.95 $\pm$ 0.17. Another fit using the same model in
RMFIT \cite{pree:2000} yields a photon index of 1.96 $\pm$ 0.18, where
the data have been rebinned to increase the S/N (see Fig. 2). In both
cases there is very good agreement between the results of SPI and the IBIS
analysis in Mereghetti et al. (2003).\nocite{mer:2003} In the energy
range 20-200\,keV the flux obtained for the burst is:
\begin{center}
$F_{\rm 20-200\,keV} = 5.5^{+5.1}_{-2.7}\, \rm erg/cm^2/s$.
\end{center}

\begin{center}
\begin{minipage}[c]{0.49\columnwidth}
\begin{center}
\includegraphics[width=4.5cm, height=6cm, angle=270]{xspec_030227}
\end{center}
\end{minipage}
\begin{minipage}[c]{0.49\columnwidth}
\begin{center}
\includegraphics[height=4.5cm, width=7cm]{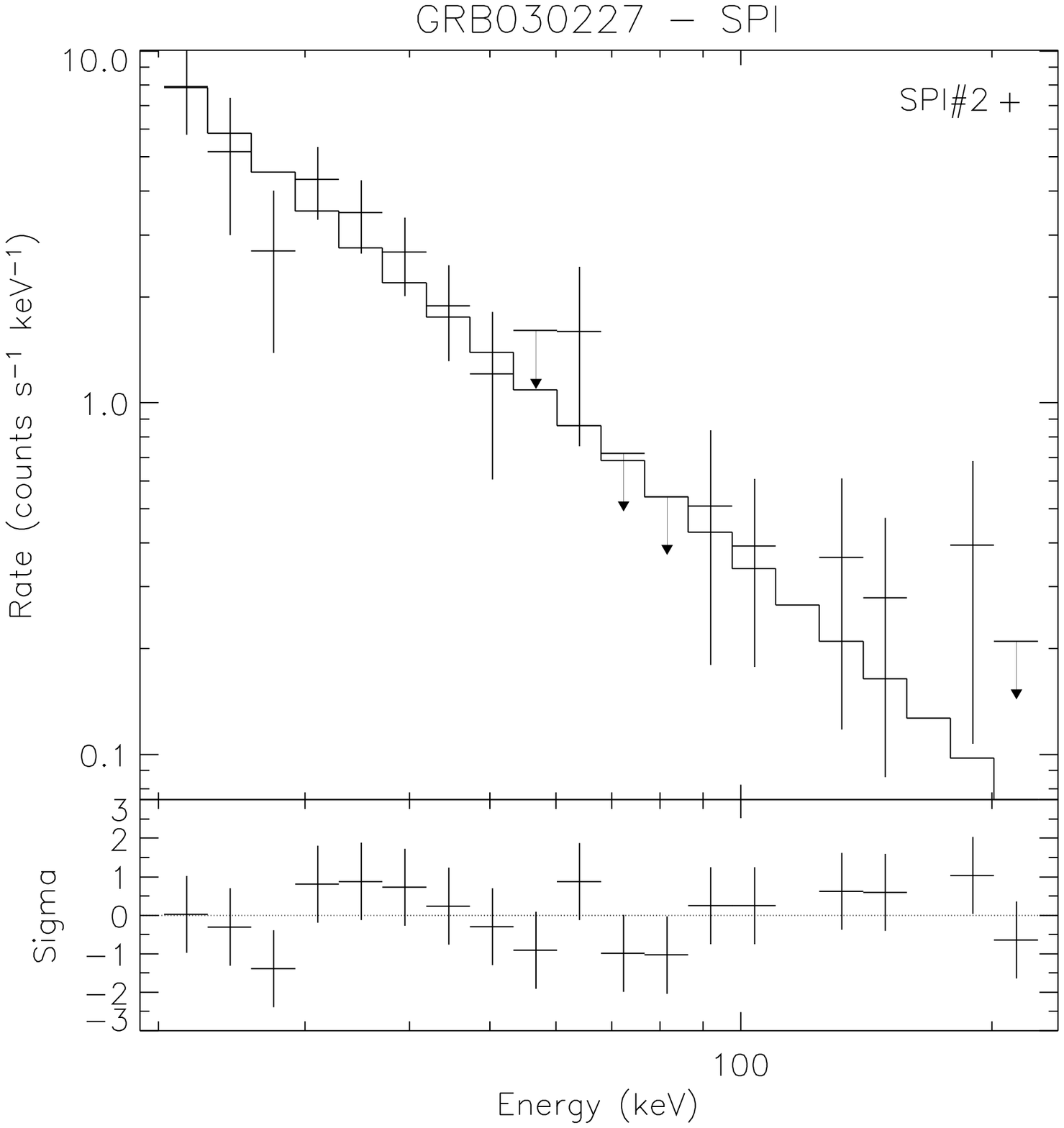}
\end{center}
\end{minipage}

\vspace{.5cm}

\small{\textbf{FIGURE 2.}\,\,\,\,\,A power law model is fit to
GRB030227 using {\it left}: XSPEC, yielding a photon index of 1.95
$\pm$ 0.17 and {\it right}: RMFIT, yielding a photon index of 1.96 $\pm$ 0.18.}

\end{center}

\vspace{-.4cm}

\section{Conclusions}

\vspace{-.2cm}

GRB030131 was the first GRB in the FOV of the main instruments after the
performance and verification phase of the satellite. The analysis of
this burst shows the capabilities of the SPI instrument during a
satellite slew, a frequent manoeuvre ($\sim$\,10\% of the time) due to
the dither pattern employed by INTEGRAL. With limited telemetry it was
still possible to image the burst and obtain a location and a
flux. Although it was not feasible to carry out spectral analysis in
this case, the imaging capability of SPI allows for cross calibration
with IBIS. For GRB030227, though a very faint burst, a spectrum can
be extracted in SPIROS and used to fit models in both XSPEC and RMFIT,
obtaining a photon index and a flux in  good agreement with previous
work \cite{mer:2003}. Therefore, SPI has demonstrated that when a
burst of sufficient intensity is observed it will be possible to study
the prompt emission of a GRB with the sensitivity necessary to
determine the spectral evolution, identify hard spectral components
and constrain models.

\vspace{-.4cm}





\small{
\bibliographystyle{aipproc}   

\bibliography{refs}

\hyphenation{Post-Script Sprin-ger}
\begin{thebibliography}{15}
\expandafter\ifx\csname natexlab\endcsname\relax\def\natexlab#1{#1}\fi
\providecommand{\enquote}[1]{``#1''}
\expandafter\ifx\csname url\endcsname\relax
  \def\url#1{\texttt{#1}}\fi
\expandafter\ifx\csname urlprefix\endcsname\relax\def\urlprefix{URL }\fi

\bibitem[Mereghetti et~al.(2002)]{mere:2003}
Mereghetti, S., et~al., \enquote{Real time localisation of Gamma Ray Bursts
  with INTEGRAL,} in \emph{Advanced Spectral Resolution}, Proceedings of the
  34th COSPAR Scientific Assembly, Houston, 2002.

\bibitem[Ubertini et~al.(2003)]{uber:2003}
Ubertini, P., et~al., \emph{A\&A}, \textbf{411}, L131--139 (2003).

\bibitem[Kn\"{o}dsleder and Roques(2002)]{knod:2002}
Kn\"{o}dsleder, J., and Roques, J.-P., \enquote{SPI Science Prospects,} in
  \emph{The Gamma-Ray Universe}, Proceedings of the XXII Moriond Astrophysics
  Meeting, Les Arcs, 2002.

\bibitem[Costa et~al.(1997)]{costa:1997}
Costa, E., et~al., \emph{Nature}, \textbf{387}, 783--785 (1997),
  [astro-ph/9706065].

\bibitem[van Paradijs et~al.(1997)]{vanp:1997}
van Paradijs, J., et~al., \emph{Nature}, \textbf{386}, 686--689 (1997).

\bibitem[Frail et~al.(1997)]{frail:1997}
Frail, D., et~al., \emph{Nature}, \textbf{389}, 261--263 (1997).

\bibitem[Metzger(1997)]{metz:1997}
Metzger, M., \emph{Nature}, \textbf{387}, 879--880 (1997).

\bibitem[Murakami et~al.(1988)]{mura:1988}
Murakami, T., et~al., \emph{Nature}, \textbf{335} (1988).

\bibitem[Briggs et~al.(1997)]{briggs:1997}
Briggs, M., et~al., \enquote{BATSE Evidence for GRB Spectral Features,} in
  \emph{Gamma-Ray Bursts}, edited by R.~P. . T.~K. C.~Meegan, Proceedings of
  the 4th Huntsville Symposium 428, AIP, New York, 1997.

\bibitem[Hjorth et~al.(2003)]{hjo:2003}
Hjorth, ., et~al., \emph{Nature}, \textbf{423}, 847--850 (2003),
  [astro-ph/0306347].

\bibitem[Gonz\'{a}lez et~al.(2003)]{gon:2003}
Gonz\'{a}lez, M., et~al., \emph{Nature}, \textbf{424}, 749--751 (2003).

\bibitem[Courvoisier et~al.(2003)]{couv:2003}
Courvoisier, T.-L., et~al., \emph{A\&A}, \textbf{411}, L53--57 (2003),
  [astro-ph/0308047].

\bibitem[{G\"{o}tz} et~al.(2003)]{got:2003}
{G\"{o}tz}, D., et~al., \emph{A\&A}, \textbf{409}, 831--834 (2003),
  [astro-ph/0307406].

\bibitem[Mereghetti et~al.(2003)]{mer:2003}
Mereghetti, S., et~al., \emph{ApJ Letters}, \textbf{590}, 73--78 (2003),
  [astro-ph/0304477].

\bibitem[Preece et~al.(2000)]{pree:2000}
Preece, R., et~al., \emph{ApJSS}, \textbf{126}, 19--36 (2000),
  [astro-ph/9908119].

\end{thebibliography}

\IfFileExists{\jobname.bbl}{}
 {\typeout{}
  \typeout{******************************************}
  \typeout{** Please run "bibtex \jobname" to optain}
  \typeout{** the bibliography and then re-run LaTeX}
  \typeout{** twice to fix the references!}
  \typeout{******************************************}
  \typeout{}
 }
}
\end{document}